\documentclass{nature}

\usepackage{amssymb}
\usepackage{amsmath}
\usepackage{wasysym}
\usepackage{graphicx}
\usepackage[usenames, dvipsnames]{color}
\usepackage[normalem]{ulem}
\usepackage{soul}

\usepackage[mathlines]{lineno}

\usepackage[numbers, sort, super]{natbib}

\usepackage{makecell}
\usepackage{hyperref}

\usepackage{subfig}
\usepackage[font=footnotesize]{caption}

\usepackage{floatrow}
\usepackage[shortcuts]{extdash}

\DeclareFloatFont{tiny}{\scriptsize}
\DeclareUnicodeCharacter{2212}{-}

\usepackage{adjustbox}
\usepackage{comment}
\usepackage{graphicx}

{\bf \title{Ultra-High-Energy Gamma-Ray Bubble around Microquasar V4641~Sgr}}

\begin{document}

\maketitle

\author{
R.~Alfaro$^{1}$,
C.~Alvarez$^{2}$,
J.C.~Arteaga-Velázquez$^{3}$,
D.~Avila Rojas$^{1}$,
H.A.~Ayala Solares$^{4}$,
R.~Babu$^{5}$,
E.~Belmont-Moreno$^{1}$,
K.S.~Caballero-Mora$^{2}$,
T.~Capistrán$^{6}$,
A.~Carramiñana$^{7}$,
S.~Casanova$^{8}$,
U.~Cotti$^{3}$,
J.~Cotzomi$^{9}$,
S.~Coutiño de León$^{10}$,
E.~De la Fuente$^{11}$,
D.~Depaoli$^{12}$,
N.~Di Lalla$^{13}$,
R.~Diaz Hernandez$^{7}$,
B.L.~Dingus$^{14}$,
M.A.~DuVernois$^{10}$,
M.~Durocher$^{14}$,
J.C.~Díaz-Vélez$^{10}$,
K.~Engel$^{15}$,
C.~Espinoza$^{1}$,
K.L.~Fan$^{15}$,
K.~Fang$^{10}$,
N.~Fraija$^{6}$,
S.~Fraija$^{6}$,
J.A.~García-González$^{16}$,
F.~Garfias$^{6}$,
A.~Gonzalez Muñoz$^{1}$,
M.M.~González$^{6}$,
J.A.~Goodman$^{15}$,
S.~Groetsch$^{5}$,
J.P.~Harding$^{14}$,
I.~Herzog$^{17}$,
J.~Hinton$^{12}$,
D.~Huang$^{15}$,
F.~Hueyotl-Zahuantitla$^{2}$,
P.~Hüntemeyer$^{5}$,
A.~Iriarte$^{6}$,
V.~Joshi$^{18}$,
S.~Kaufmann$^{19}$,
D.~Kieda$^{20}$,
C.~de León$^{3}$,
J.~Lee$^{21}$,
H.~León Vargas$^{1}$,
J.T.~Linnemann$^{17}$,
A.L.~Longinotti$^{6}$,
G.~Luis-Raya$^{19}$,
K.~Malone$^{22}$,
O.~Martinez$^{9}$,
J.~Martínez-Castro$^{23}$,
J.A.~Matthews$^{24}$,
P.~Miranda-Romagnoli$^{25}$,
J.A.~Morales-Soto$^{3}$,
E.~Moreno$^{9}$,
M.~Mostafá$^{26}$,
A.~Nayerhoda$^{8}$,
L.~Nellen$^{27}$,
M.~Newbold$^{20}$,
M.U.~Nisa$^{17}$,
R.~Noriega-Papaqui$^{25}$,
L.~Olivera-Nieto$^{12}$,
N.~Omodei$^{13}$,
M.~Osorio$^{6}$,
Y.~Pérez Araujo$^{1}$,
E.G.~Pérez-Pérez$^{19}$,
C.D.~Rho$^{28}$,
D.~Rosa-González$^{7}$,
E.~Ruiz-Velasco$^{12}$,
H.~Salazar$^{9}$,
D.~Salazar-Gallegos$^{17}$,
A.~Sandoval$^{1}$,
M.~Schneider$^{15}$,
J.~Serna-Franco$^{1}$,
A.J.~Smith$^{15}$,
Y.~Son$^{21}$,
R.W.~Springer$^{20}$,
O.~Tibolla$^{19}$,
K.~Tollefson$^{17}$,
I.~Torres$^{7}$,
R.~Torres-Escobedo$^{29}$,
R.~Turner$^{5}$,
F.~Ureña-Mena$^{7}$,
E.~Varela$^{9}$,
L.~Villaseñor$^{9}$,
X.~Wang$^{5}$,
I.J.~Watson$^{21}$,
E.~Willox$^{15}$,
S.~Yun-Cárcamo$^{15}$,
H.~Zhou$^{29}$
}

\begin{affiliations}
\small
  \item Instituto de F\'{i}sica, Universidad Nacional Autónoma de México, Ciudad de Mexico, Mexico
  \item Universidad Autónoma de Chiapas, Tuxtla Gutiérrez, Chiapas, México
  \item Universidad Michoacana de San Nicolás de Hidalgo, Morelia, Mexico 
  \item Department of Physics, Pennsylvania State University, University Park, PA, USA 
  \item Department of Physics, Michigan Technological University, Houghton, MI, USA 
  \item Instituto de Astronom\'{i}a, Universidad Nacional Autónoma de México, Ciudad de Mexico, Mexico 
  \item Instituto Nacional de Astrof\'{i}sica, Óptica y Electrónica, Puebla, Mexico
  \item Institute of Nuclear Physics Polish Academy of Sciences, PL-31342 IFJ-PAN, Krakow, Poland 
  \item Facultad de Ciencias F\'{i}sico Matemáticas, Benemérita Universidad Autónoma de Puebla, Puebla, Mexico
  \item Department of Physics, University of Wisconsin-Madison, Madison, WI, USA 
  \item Departamento de F\'{i}sica, Centro Universitario de Ciencias Exactase Ingenierias, Universidad de Guadalajara, Guadalajara, Mexico
  \item Max-Planck Institute for Nuclear Physics, 69117 Heidelberg, Germany
  \item Department of Physics, Stanford University: Stanford, CA 94305–4060, USA
  \item Physics Division, Los Alamos National Laboratory, Los Alamos, NM, USA 
  \item Department of Physics, University of Maryland, College Park, MD, USA 
  \item Tecnologico de Monterrey, Escuela de Ingenier\'{i}a y Ciencias, Ave. Eugenio Garza Sada 2501, Monterrey, N.L., Mexico, 64849
  \item Department of Physics and Astronomy, Michigan State University, East Lansing, MI, USA 
  \item Erlangen Centre for Astroparticle Physics, Friedrich-Alexander-Universit\"at Erlangen-N\"urnberg, Erlangen, Germany
  \item Universidad Politecnica de Pachuca, Pachuca, Hgo, Mexico 
  \item Department of Physics and Astronomy, University of Utah, Salt Lake City, UT, USA 
  \item University of Seoul, Seoul, Rep. of Korea
  \item Space Science and Applications Group, Los Alamos National Laboratory, Los Alamos, NM, USA
  \item Centro de Investigaci\'on en Computaci\'on, Instituto Polit\'ecnico Nacional, M\'exico City, M\'exico.
  \item Dept of Physics and Astronomy, University of New Mexico, Albuquerque, NM, USA 
  \item Universidad Autónoma del Estado de Hidalgo, Pachuca, Mexico 
  \item Department of Physics, Temple University, Philadelphia, PA, USA
  \item Instituto de Ciencias Nucleares, Universidad Nacional Autónoma de Mexico, Ciudad de Mexico, Mexico 
  \item Department of Physics, Sungkyunkwan University, Suwon 16419, South Korea
  \item Tsung-Dao Lee Institute, Shanghai Jiao Tong University, Shanghai, China
\end{affiliations}

{\bf Microquasars are laboratories for the study of jets of relativistic particles produced by accretion onto a spinning black hole.  Microquasars are near enough to allow detailed imaging of spatial features across the multiwavelength spectrum.  The recent extension of the spatial morphology of a microquasar, SS~433, to TeV gamma rays \cite{abeysekara2018very} localizes the acceleration of electrons at shocks in the jet far from the black hole \cite{hess2024ss433}.  Here we report TeV gamma-ray emission from another microquasar, V4641~Sgr, which reveals particle acceleration at similar distances from the black hole as SS~433.  Additionally, the gamma-ray spectrum of V4641 is among the hardest TeV spectra observed from any known gamma-ray source and is detected up to 200 TeV. Gamma rays are produced by particles, either electrons or hadrons, of higher energies. Because electrons lose energy more quickly the higher their energy, such a spectrum either very strongly constrains the electron production mechanism or points to the acceleration of high-energy hadrons.  
This observation suggests that large-scale jets from microquasars could be more common than previously expected and that microquasars could be a significant source of Galactic cosmic rays. 
high energy gamma-rays also provide unique constraints on the acceleration mechanisms of extra-Galactic cosmic rays postulated to be produced by the supermassive black holes and relativistic jets of quasars. The distance to quasars limits imaging studies due to insufficient angular resolution of gamma-rays and due to attenuation of the highest energy gamma-rays by the extragalactic background light.}

Previous observations of V4641~Sgr did not report gamma-ray emission \cite{abdalla2018search}. The microquasar V4641 Sagittarii (V4641~Sgr) is a binary system with a black hole and main sequence B-type companion star~\cite{MacDonald_2014, gaia2018gaia}. V4641 Sgr stands out for its super-Eddington accretion \cite{1999IAUC.7119....1I} and for its radio jet, which is one of the fastest superluminal jets in the Milky Way. Observations of this region from 2015 to 2022 by the HAWC Observatory~\cite{NIM2023} have revealed significant gamma-ray emission coincident with the location of V4641~Sgr, as shown in Figure~\ref{fig:map}. The excess over the estimated cosmic-ray background flux reaches a maximum significance of 8.8$\sigma$ above 1~TeV and 5.2$\sigma$ above 100~TeV. After ruling out an association with extragalactic background sources and other high-energy sources in the Galaxy, we suggest V4641~Sgr as the likely source of the observed gamma-ray excess (see details in ~\hyperref[sec:associations]{Methods}).

Based on a systematic multi-source analysis method on a 3$^{\circ}$ radius region of interest~(ROI) around the gamma-ray emission, the excess may be either described as two point-like sources or one extended source with an asymmetric Gaussian distribution (see details in ~\hyperref[sec:modeling]{Methods}). 
The current statistics do not allow us to distinguish between the two spatial models. When adopting a two-point-source model, the northern source and southern source are detected at 8.1$\sigma$ and 6.7$\sigma$, respectively, with best-fit locations situated approximately  30 and 60~pc (considering a distance of 6.6~kpc) away from the binary system, respectively.
A single-point-source template is disfavored at 8.3$\sigma$ with respect to an asymmetric extended-source template, strongly suggesting that the gamma-ray emission is from a region more extended than the central binary.

\begin{figure}[tpb]
\begin{center}
\resizebox{0.6\textwidth}{!}{
\includegraphics[width=0.7\textwidth]{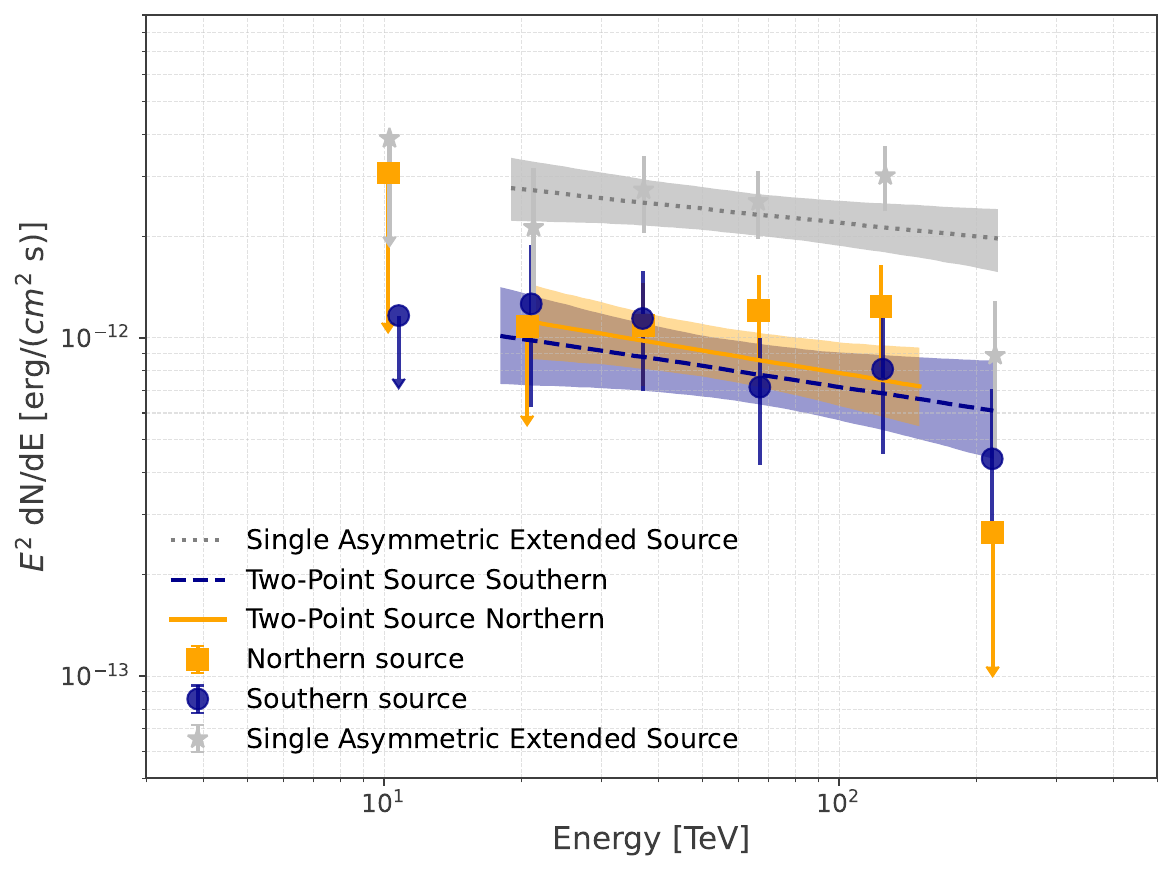}

}
\caption{ 
{Differential spectrum weighted by E$^2$  for the northern and southern sources in a model with two point sources, and for the asymmetric extended source in a model with a single asymmetric extended source.} The shaded regions indicate the best-fit spectra and $1\sigma$ statistical uncertainties when fitting a single power-law model to the data from 10 to \textgreater 200~TeV. The markers correspond to the best-fit values and their $1\sigma$ statistical uncertainties obtained when fitting a single-power-law model to data in individual energy bins. The chosen energy range for plotting the spectrum is specified in~\hyperref[sec:energyrange]{Methods}. }

\label{fig:spectra}
\end{center}
\end{figure}

Considering the model with two point sources from a physically grounded perspective, the excess spectrum extends up to 217~TeV without any observable indication of a cutoff (for best-fit results of the model with a single asymmetric extended source, see ~\hyperref[sec:modeling]{Methods})).  
The spectrum above 1~TeV
is best described by a power law, $dN/dE=N_{0}(E/E_{0})^{\Gamma}$, where $E_{0}=47~\mathrm{TeV}$ is the pivot energy. The pivot energy is selected to minimize the correlation between parameters due to the choice of the spectral model. Table~\ref{tab:2ps-2pctfHit-result} lists the best-fit values of $N_0$ and $\Gamma$ for the northern and southern sources. We note that the exceptionally hard spectrum, $\Gamma = - 2.2$, makes V4641~Sgr among the hardest ultra-high-energy (UHE) sources ever measured.  Figure~\ref{fig:spectra} compares the spectra of the two sources. Despite being $\sim$0.69$^{\circ} \pm 0.04$ ($= 80~\rm pc$) 
away from each other, the two sources present almost identical flux amplitudes and spectral indices, hinting that they likely share a common origin. 
Considering that no other plausible multiwavelength counterparts can be identified and that the two point-like sources present remarkably similar spectra while being physically distant,   
the origin of the HAWC excess is likely connected to V4641~Sgr and could be due to persistent large-scale outflows from the system, which we refer to as a bubble.

Interaction of large-scale jets with the interstellar medium (ISM) may induce high-energy radiation. So far, SS~433 is the only microquasar with very-high-energy (VHE; 0.1--100~TeV) gamma-ray emission observed from the lobes \cite{abeysekara2018very, hess2024ss433}.  
Assuming a distance of 6.6~kpc for V4641~Sgr \cite{gaia2018gaia}, the physical separation between each of the two sources and the central object is at the level of tens of parsecs. Our observation implies that V4641~Sgr could be closely analogous to SS~433 \cite{abeysekara2018very}, which has long been proposed to be based on optical and X-ray observations of the flares \cite{2002A&A...385..904R,lindstrom2005new,gallo2013v4641}. The ratio of the TeV gamma-ray power and the Eddington luminosity of this source is an order of magnitude higher than that of SS~433, suggesting that large-scale outflows from microquasars may carry high kinetic power and be efficient particle accelerators.

Persistent VHE gamma-ray emission from microquasars can be expected from accelerated electrons inverse Compton scattering off low-energy photons (leptonic scenario) or from the decay of neutral pions, which are produced by the interaction of protons and nuclei (hadronic scenario) \cite{aharonian1998gamma, 2002A&A...390..751H}.

A leptonic scenario is challenging for the following reasons. Firstly, a fast outflow is needed to accelerate electrons to 200~TeV and above. The acceleration time, $t_{acc} \sim 10 D(E_e)/{v_{sh}}^2~\mathrm{yr}$, needs to be shorter than the cooling time due to synchrotron radiation in a magnetic field $B$, $t_{\rm cooling}=t_{\rm synch} \sim 500 {(\frac{E_e}{200 \, \rm TeV})}^{-1} {(\frac{B}{10 \,\mu G})}^{-2}~\mathrm{yr}$, yielding a shock velocity, $v_{\rm sh}>7000 \mathrm{km/s}$, equivalent to 2\% of the speed of light. Here, $D(E_e) = \eta R_L \,c /3$ is the Bohm diffusion coefficient and $R_L$ is the Larmor radius of the particle. Secondly, electrons at such high energies cool so quickly that they can hardly travel over 100~pc.  The diffusion time, $R^2/(2D) \sim 1000/\eta~\mathrm{yr}$, where $R\sim 100~\mathrm{pc}$, $D(200~\mathrm{TeV})\sim \eta 10^{30}~\mathrm{cm}^2/\mathrm{s}$ and $\eta  \ll 1$, is much longer than the cooling time.

\begin{figure}[htpb]
\begin{center}
\resizebox{0.6\textwidth}{!}{
\includegraphics[width=0.7\textwidth]{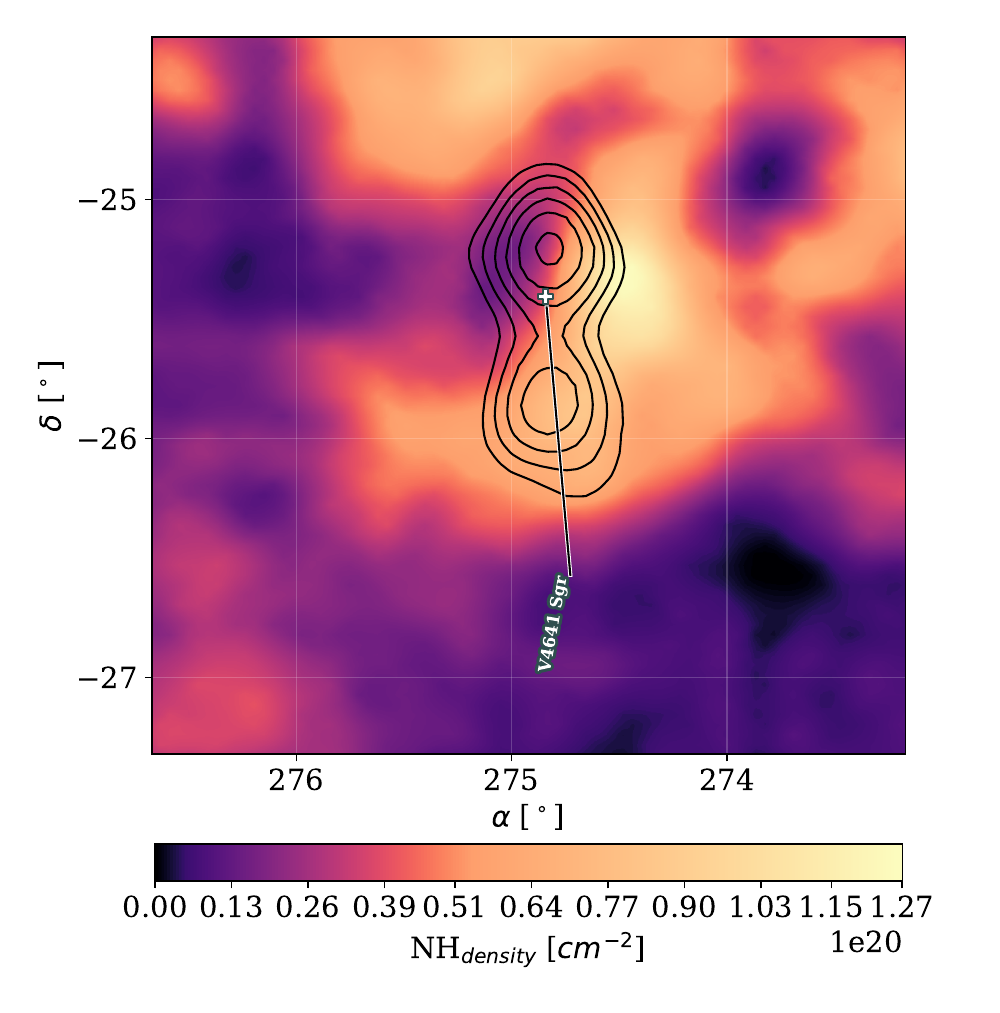}
}
\caption{
Gas distribution at the location of V4641~Sgr. The black contours represent the VHE gamma-ray excess obtained by HAWC. The column density of atomic hydrogen, $N_H$, at the V4641~Sgr location is obtained by integrating the gas survey cubes over the range 70--120~km/s and then dividing by the velocity interval. The level of noise in the molecular hydrogen map is too high to confirm or rule out the presence of the target material in molecular form.
}
\label{fig:gas}
\end{center}
\end{figure}

In the hadronic scenario, protons are accelerated to PeV energies and interact with the ambient gas, producing neutral pions that quickly decay into gamma rays. Extended Data Figure~\ref{fig:gas} in the \hyperref[sec:gas]{Materials} section shows the gas distribution near the gamma-ray excess detected by HAWC. To account for both the southern and northern HAWC sources, we require 
a total proton energy $W_p \sim {1} \times {10}^{50}~\mathrm{erg}$ for these two sources.
The protons could be accelerated at the termination shock, where the jets impact with the ISM, or along the jets and subsequently transported to the HAWC sources. 
We assume each HAWC source has a radius of 20~pc, which corresponds to the upper limit on the source radius of 0.2~degrees at 95\% confidence level, and consider escape due to diffusion for two cases: diffusion as inferred at GeV energies from the cosmic-ray secondaries, or the much slower Bohm diffusion. For 1~PeV protons, the escape time is then 40~yrs or 4,000~yrs, respectively. The energy required in the former case exceeds the Eddington luminosity assuming protons are accelerated with the same spectrum from 1~GeV to 1~PeV, but is only a small fraction of the Eddington luminosity in the latter case. We also note that the power channeled into the kinetic energy of the jets might easily exceed the Eddington luminosity, as testified by the super-Eddington flares \cite{2004vhec.book.....A}.

The detection of greater than 100~TeV photons indicates that microquasars could be protonic PeVatrons. 
A smoking gun for this hadronic Pevatron scenario could be the detection of high-energy neutrinos from V4641~Sgr. 
Subsequent multi-wavelength and multi-messenger observations centered on the VHE gamma-ray emission site will yield improved constraints on the magnetic-field intensity and the properties of the parent-particle species and population.

Large-scale jets have been found in a few microquasars \cite{1984ARA&A..22..507M, doi:10.1126/science.1075857, 2005Natur.436..819G,  2010ApJ...719L.194S, 2010Natur.466..209P}. They carry the bulk of the liberated accretion power of the compact objects and are local analogs of Fanaroff–Riley type~II active galaxies. The interaction of large-scale jets and the ambient medium is suggested to produce shock-ionized nebulae like W50 \cite{2004ASPRv..12....1F} and to explain the bubbles around ultraluminous X-ray sources, the most luminous class of extra-nuclear X-ray emitters \cite{2003RMxAC..15..197P, 2020ApJ...896..117B, 2010Natur.466..209P,  2017ARA&A..55..303K}. 
The formation of large-scale jets and their connection with extended nebulae or bubbles around microquasars remains poorly known because of the rarity of such objects.   
Detection of extended jets is challenging, both because the outflows could be radiatively inefficient and thus ``dark" in optical/UV/X-ray \cite{2005Natur.436..819G} and because a pair of extragalactic sources located on either side of a microquasar could cause a false association in radio and X-ray wavelengths \cite{1999ARA&A..37..409M}. 
  
With a jet luminosity exceeding $L_{\rm edd} \sim 10^{39}\,\rm erg\,s^{-1}$ \cite{2002A&A...385..904R}, the gamma-ray excess observed by HAWC aligns with the expectation that V4641~Sgr powers a wind-driven nebula or jet-driven radio lobes at a scale of $R\approx 0.76 (L_{\rm jet} / \rho_0)^{1/5} \,t^{3/5} = 110~ \rm pc$, assuming an ambient mass density $\rho_0 \sim   (1\,{\rm cm}^{-3}) \, m_p$, with $m_p$ being the mass of the proton, and age $t\sim 1\,\rm Myr$ \cite{2020MNRAS.495.2179S}. 
The discovery of the UHE gamma-ray bubble around V4641~Sgr supports the longstanding hypothesis that this source shares similarities with SS~433 and suggests that the super-Eddington outflows could power steady, large-scale jets. 

Such jets may play an important role in the production of Galactic cosmic rays. Microquasar jets have long been suggested as cosmic-ray sources, though the level of their contribution to the Galactic cosmic-ray flux is largely unknown. As a result of the uncertainties in the duty cycle, kinetic energy output, and composition of the microquasar jets, this contribution may vary from 0.1\% to 100\% \cite{Heinz:2002qj,Cooper:2020tzq}. Our observation favors microqusar jets as extraordinary cosmic-ray sources by suggesting the existence of large-scale steady jets, thereby having a long duty cycle, and by finding V4641~Sgr to be an emitter of 200~TeV photons, such that it is, therefore, a plausible accelerator of PeV cosmic-ray protons.

\clearpage

\begin{figure}[ht!]
\begin{center}
\resizebox{1.0\textwidth}{!}{%
\includegraphics[width=0.5\textwidth]{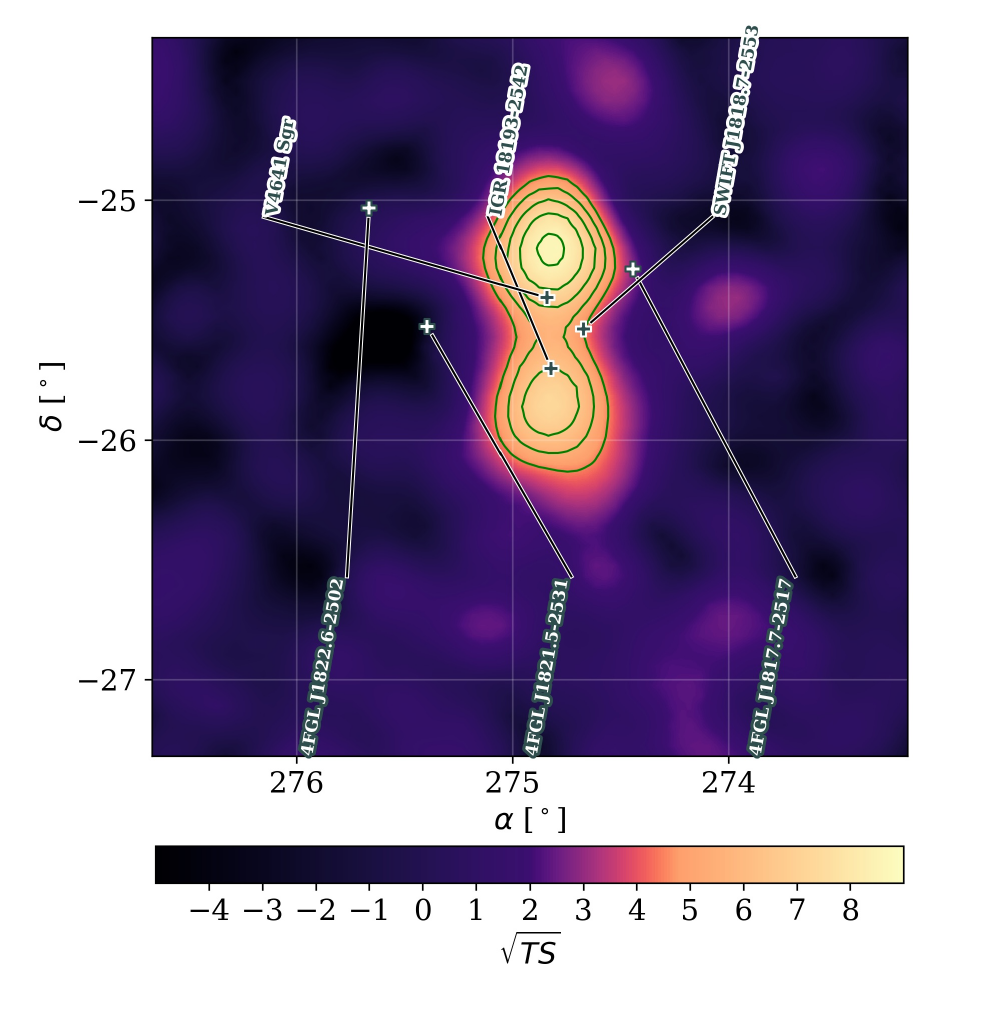}%
\quad
\includegraphics[width=0.5\textwidth]{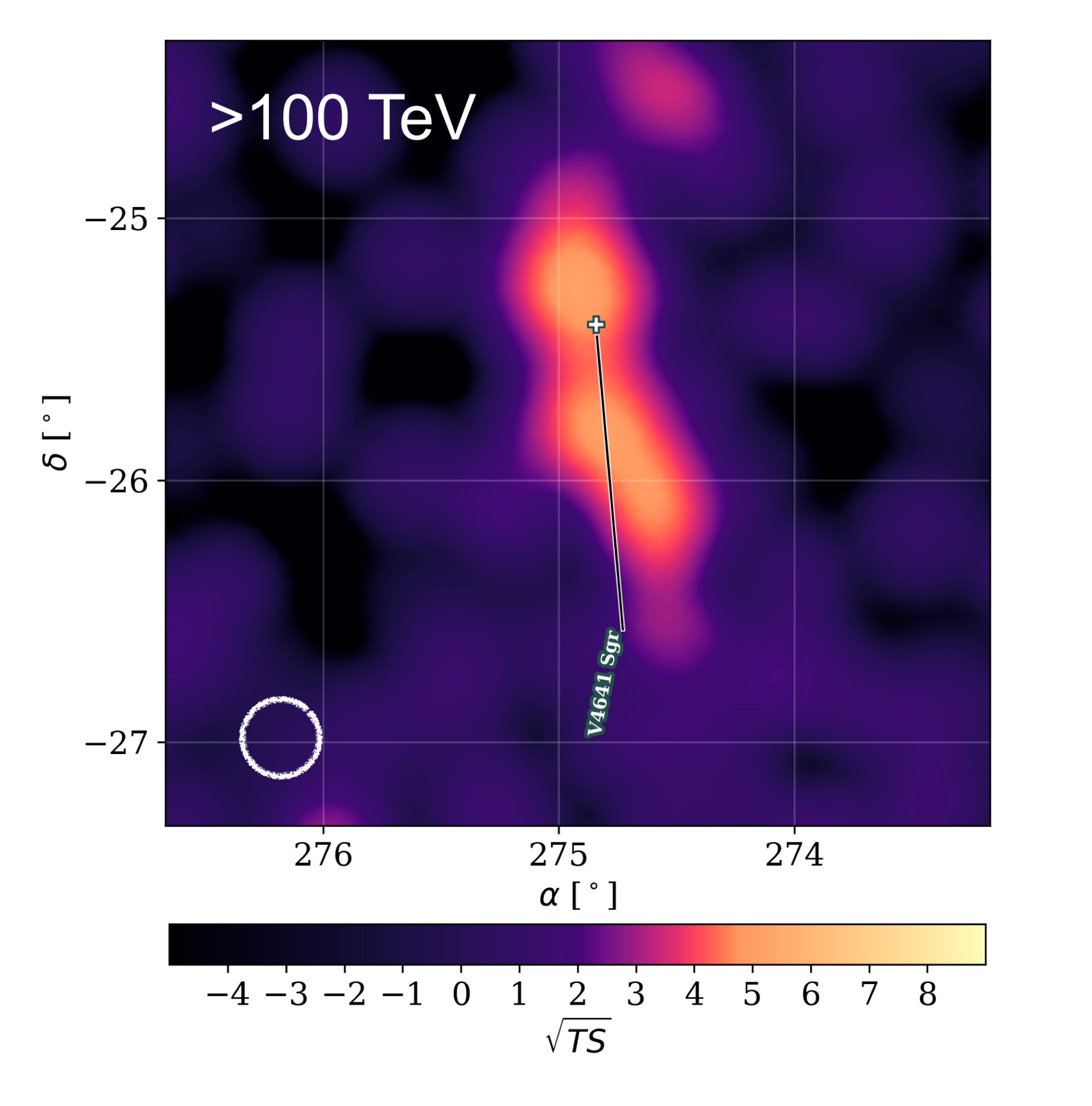}%
}
\caption{Significance map around the V4641 Sgr region for events with median energy greater than 1~TeV. The value TS refers to the likelihood ratio test statistic described in Equation~\ref{ts} in the~\hyperref[sec:methods]{Methods} section. The green contours indicating significance are mapped to $\rm{\sqrt{TS}}$ values ranging from 4.5 to 8.5, increasing inwards at intervals of one from the outermost contour to the innermost. The black stars represent the best-fit locations from the two-point-source model; \textbf{(b)}: Significance map (of the same region) including only events with reconstructed energies greater than 100~TeV. The white circle represents the angular resolution at a radius corresponding to 68\% event containment (0.17$^\circ$) at this energy range. The V4641 Sgr location is taken from Ref.~\citenum{gaia2018gaia}. These significance maps are made by assuming a point-source hypothesis and a power-law spectrum with the best-fit index $-2.2$.}
    
\label{fig: map}
\end{center}
\end{figure}

\begin{table}[tpb]
\renewcommand{\arraystretch}{2}
  \centering
  \begin{adjustbox}{max width=\textwidth}
    \begin{tabular}{ |c|c|c|c|c|c|c|} 
      \hline
      Source Name & R.A. [$^{\circ}$] & Dec. [$^{\circ}$]  & $\phi_{0}$ [$\times 10^{-16}\rm cm^{-2} TeV^{-1} s^{-1}$] & Index ($\alpha$)  & \makecell{ { Extension upper limit} \\ { at 95\% CL [$^{\circ}$]}} & \makecell{{ Physical size} \\ {(distance: 6.6 kpc)}}\\
      \hline
      Southern & 274.82 $\pm 0.04$ & -25.87 $\pm 0.03$  & $2.4^{+0.6}_{-0.5}(stat.)^{+ 0.2}_{-0.5}(syst.)$ & $-2.2 \pm 0.2 (stat.)^{+ 0.07}_{-0.02}(syst.)$ & {0.2} & $\sim$ {30~pc}\\ 
      \hline
      Northern & 274.82$\pm 0.03$ & -25.18 $\pm 0.02$  & $2.6^{+0.5}_{-0.4}(stat.) \pm 0.4(syst.)$ & $-2.2 \pm 0.2 (stat.)^{+ 0.07}_{-0.05}(syst.)$ & {0.2} & { $\sim$ 60~pc}\\
      \hline
      
    \end{tabular}
  \caption{Best-fit parameters for a model with two point sources; the optimal pivot energy, $E_0$, is 47~TeV for both sources.  }
  \label{tab:2ps-2pctfHit-result}

\end{adjustbox}
\end{table}


\clearpage
\newpage

\subsection{Acknowledgements} We acknowledge the support from: the US National Science Foundation (NSF); the US Department of Energy Office of High-Energy Physics; the Laboratory Directed Research and Development (LDRD) program of Los Alamos National Laboratory; Consejo Nacional de Ciencia y Tecnolog\'{i}a (CONACyT), M\'{e}xico, grants 271051, 232656, 260378, 179588, 254964, 258865, 243290, 132197, A1-S-46288, A1-S-22784, CF-2023-I-645, c\'{a}tedras 873, 1563, 341, 323, Red HAWC, M\'{e}xico; DGAPA-UNAM grants IG101323, IN111716-3, IN111419, IA102019, IN106521, IN110621, IN110521 , IN102223; VIEP-BUAP; PIFI 2012, 2013, PROFOCIE 2014, 2015; the University of Wisconsin Alumni Research Foundation; the Institute of Geophysics, Planetary Physics, and Signatures at Los Alamos National Laboratory; Polish Science Centre grant, DEC-2017/27/B/ST9/02272; Coordinaci\'{o}n de la Investigaci\'{o}n Cient\'{i}fica de la Universidad Michoacana; Royal Society - Newton Advanced Fellowship 180385; Generalitat Valenciana, grant CIDEGENT/2018/034; The Program Management Unit for Human Resources \& Institutional Development, Research and Innovation, NXPO (grant number B16F630069); Coordinaci\'{o}n General Acad\'{e}mica e Innovaci\'{o}n (CGAI-UdeG), PRODEP-SEP UDG-CA-499; Institute of Cosmic Ray Research (ICRR), University of Tokyo, National Research Foundation of Korea (RS-2023-00280210). H.F. acknowledges support by NASA under award number 80GSFC21M0002. We also acknowledge the significant contributions over many years of Stefan Westerhoff, Gaurang Yodh, and Arnulfo Zepeda Dominguez, all deceased members of the HAWC collaboration. Thanks to Scott Delay, Luciano D\'{i}az, and Eduardo Murrieta for technical support.

\subsection{Auhor contributions} 
X.~Wang (xwang32@mtu.edu) and D.~Huang (dezhih@umd.edu) analyzed the data and performed the maximum likelihood analysis. Physics modeling and interpolation was carried out by K. Fang (kefang@physics.wisc.edu) and S.~Casanova (sabrina.casanova@ifj.edu.pl). B.~Dingus (bldingus@gmail.com), J.~A.~Goodman (goodman@umd.edu), and P.~Huntemeyer (petra@mtu.edu) helped to improve the manuscript. The full HAWC Collaboration has contributed to the construction, calibration, and operation of the detector, the development and maintenance of reconstruction and analysis software, and the vetting of the analysis presented in this manuscript. All authors have reviewed, discussed, and commented on the results and the manuscript.

\subsection{Competing interests}
The authors declare no competing interests.

\newpage

\section*{Methods}
\label{sec:methods}


\subsection{HAWC data}
\label{sec:observation}
~The HAWC Observatory 
consists of 300 water Cherenkov detectors located on the slope of the Sierra Negra volcano in Puebla, Mexico, at an elevation of 4,100 meters \cite{NIM2023}. Each detector has four upward-facing photomultiplier tubes (PMTs) at its bottom to collect the Cherenkov light produced by secondary air-shower particles in the water. Using the charge and time information recorded by the PMTs, we can reconstruct the properties of the primary particles. Recently, with better reconstruction algorithms and gamma/hadron separation, HAWC's sensitivity has been improved by a factor of 2--4 (depending on the declination and energy range). 

This analysis uses $\sim$2,400 days of data collected between 26th November 2014 and 27th June, 2022, reconstructed using recently improved algorithms{ \cite{albert2024performance}} The HAWC data are binned according to the fraction of available PMTs hit and reconstructed energy \cite{crab2019}; air-shower events with higher energy typically trigger more PMTs. At the same time, we also distinguish the events whose shower cores land exclusively on the array (on-array events) from those with cores landing off the array (off-array events). Only on-array events are used in this work. Events with the highest fraction of PMTs hit have the highest energies and the best angular resolutions, with values $0.18^{\circ}$ (68\% containment radius) or better.

\subsection{Likelihood analysis}
\label{sec:modeling}
We performed a maximum-likelihood analysis using the HAWC plugin to the Multi-Mission Maximum Likelihood (3ML) software framework~\cite{3ml,younk2015high,chadHAL2021}. This method obtains the best-fit parameters via maximum likelihood for a model with a given spatial and spectral assumption convolved with the detector response functions. The background is estimated by the ``direct integration" method~\cite{atkins2003observation,crab2017}. For events with a larger fraction of PMT hits, we also used the ``background randomization"~\cite{crab2019} method as a complementary method with ``direct integration" to spatially smooth the background.

The likelihood ratio test statistic (TS) is used to compare the two given models, defined as:
\begin{equation}\label{ts}
    TS=-2~\ln\bigg(\frac{L_{\textbf{null}}}{L_{\text{alt}}}\bigg)\, ,
\end{equation}
where $L_{\text{alt}}$ is the maximum likelihood of the alternative hypothesis~(background + source model) and $L_{\text{null}}$ is the likelihood of the null hypothesis~(background only).

In the analysis, we used a circular ROI with a radius of 3$^\circ$, centered at $R.A.=274.92^{\circ}$, $Dec.=-25.82^{\circ}$.  
To explore gamma-ray emission within the ROI, we conducted a systematic multi-source analysis method which is inspired by the \textit{Fermi}-LAT \cite{Fermi-LAT-extended}. 
As the latitude of the center of the ROI is $5^\circ$ away from the Galactic plane, background emission due to the Galactic diffuse emission and unresolved sources on the Galactic plane is negligible. 
Additionally, because the source is so isolated, unlike SS 433 and other sources on the galactic plane that could be affected by galactic diffuse emission, it serves as an excellent laboratory for studying the source emission independently. Therefore, we begin with a model consisting of a single point source and a simple-power-law spectral assumption. Then, we add point sources until there is no excess with significance greater than 4$\sigma$. Following this, we test the extension of each added point source sequentially to obtain the best morphological description of the emission. We further explore two distinct spectral models for the emission: the simple-power-law spectrum ($dN/dE=N_{0}(E/E_{0})^\alpha$) and the log-parabola spectrum ($dN/dE=N_{0}(E/E_{0})^{\alpha-\beta\log(E/E_{0})}$).  The pivot energy, $E_{0}$, is determined by minimizing the correlation between the flux normalization, $N_{0}$, and the spectral index, $\alpha$. The pivot energy is fixed at 47~TeV for models with two sources while letting the other parameters float.
Through this comprehensive multi-source analysis, we identified a model with two point sources as the best model to describe the observed gamma-ray excess. Since this analysis is for a targeted, specific region, we further explored several alternative models involving asymmetric extended sources. A single asymmetric extended source model also provides a satisfactory description of our data.


\begin{table}[htpb]
\renewcommand{\arraystretch}{2}
\begin{center}
\begin{tabular} { |c|c|c|c|}
\hline
Model & $-$logLikelihood & BIC & AIC \\ \hline
One Point Source & 60733 & 121520 & 121473 \\ \hline
One Asymmetric Extended Source & 60694 & 121485 & 121403 \\ \hline
Two Point Sources & 60694  &  121498 & 121404 \\ \hline
\end{tabular}
\end{center}
\caption{Comparison of the performance of different models. }
\label{tab:model_comaparisons}
\end{table}

The value $\Delta$TS is used to compare nested models, such as the single-point-source model and the model with two point sources. The Bayesian information criterion (BIC~\cite{kass1995bayes,liddle2007information}) and Akaike information criterion (AIC~\cite{AIC}) are used for non-nested model comparisons. 
In the model with two point sources, the $\Delta$TS between the simple-power-law spectrum and the log-parabola spectrum is 0 for the southern source and 6 for the northern source; for the model with a single asymmetric extended source, the $\Delta$TS between the two spectral models is 6.
Given that the values are 0 and 2.4$\sigma$, respectively, for one additional free parameter, both of which being \textless 4$\sigma$, this comparison suggests that there is no significant curvature in the energy spectrum within the current HAWC data.
The upper limit at 95\% confidence level for the Gaussian width of the sources in the two-source model is $0.23^{\circ}$ and $0.17^{\circ}$ for the southern source and northern source respectively.

\begin{table}[tpb]
\scriptsize
\renewcommand{\arraystretch}{2}
  \begin{center}
    \begin{tabular} { |c|c|c|c|c|c| } 
      \hline
       R.A. [$^{\circ}$] & Dec. [$^{\circ}$]  & { Gaussian Width of Major Axis} [$^{\circ}$] & $\phi_{0}$ [$\times10^{-16}\rm cm^{-2} TeV^{-1} s^{-1}$] & Index & Pivot Energy [TeV] \\
      \hline
       $274.81 \pm 0.03$ & $-25.56 \pm 0.09$  & $0.54 \pm 0.08$ & $4.5^{+0.7}_{-0.6}$ & $-2.1 \pm 0.1 $ & $57$\\ 
      \hline    
    \end{tabular}
  \end{center}
  \caption{Best-fit parameters for the asymmetric extended-source model }
  \label{tab:1asym-result}
\end{table}

 From Table \ref{tab:model_comaparisons}, it is evident that the single-point-source model is disfavored in comparison to the model with two point sources, with a $\Delta$TS value of 78, corresponding to an 8.1$\sigma$ significance, considering four degrees of freedom. Additionally, the comparison between a single symmetric extended source model and a model with a single asymmetric extended source is unfavorable based on both $\Delta$TS and $\Delta$BIC assessments. It is noteworthy that the $\Delta$BIC value incurs significant penalties due to the two extra parameters and the large number of pixels in the ROI associated with the asymmetric extended source model. Moreover, the current HAWC data does not allow us to differentiate between the model with two point sources and the model with a single asymmetric extended source. The presence of an extra free parameter and the involvement of nearly 100,000 pixels in the fitting process naturally leads to a substantial $\Delta$BIC, disfavoring the model with two point sources. If we disregard the penalty originating from the pixel count in the fit and consider instead the AIC, the model with a single asymmetric extended source only exhibits a marginal improvement with a $\Delta$AIC value of one. This small difference emphasizes the inability to distinguish between the two models. Given the best-fit eccentricity of $0.98 \pm 0.01$ (for other parameters see Table~\ref{tab:1asym-result}), it remains plausible that the occurrence of the asymmetric extended source could be attributed to the presence of two closely overlapping sources that cannot be definitively distinguished. We also tested models fixed at the black hole location for events above 56~TeV. Compared to the model with a single extended source, a single point source is disfavored with $\Delta\mathrm{TS}=34$, equivalent to 5.9$\sigma$ with one degree of freedom.

\subsection{Energy-range study/upper limit on the observed photon energy}
\label{sec:energyrange}
To estimate the maximum energy of the gamma-ray emission detected by the HAWC Observatory, we applied a forward-folding method similar to { Abeysekara, A. \emph{et al.} }~\cite {geminga2017,binata2021hawc}. In this method, we multiply a step function by the best-fit energy spectral model (a simple-power-law spectrum). 
Keeping the boundary of the function floating and all the other parameters fixed in the fitting, we can obtain the maximum and minimum energy by comparing the log-likelihood value with our best-fit model.
The energy value at which the log-likelihood value decreases by $1\sigma$ is chosen as the minimum $E_{max}$ energy for the detected gamma rays. Based on the energy-range study, the current measurement of the gamma-ray excess observed by HAWC extends from 18~TeV to 217~TeV for the southern source, while the northern source is observed from 21~TeV to 150~TeV. Both sources share identical spectral indices of $-2.20$, as shown in Figure~\ref{fig:spectra}.

\subsection{Systematic uncertainties}
\label{sec:systematic_errors}
The contribution to the systematic uncertainties from the detector effects on the flux normalization of the TeV emission from V4641~Sgr is less than $\pm20\%$ of the nominal value. The index of the TeV emission from V4641~Sgr changes by \textless$3\%$ because of detector systematics. These systematic uncertainties
are determined as described in Ref.~\citenum{crab2019}, which includes the absolute quantum efficiency of the PMTs, charge resolution and threshold of the PMTs, 
changes to PMT efficiency over time, and how late the light in the air shower is treated. Uncertainties from each category are investigated and summed in quadrature to estimate the total systemic uncertainties.

\begin{figure}[htpb]
\begin{center}
\resizebox{0.6\textwidth}{!}{
\includegraphics[width=0.9\textwidth]{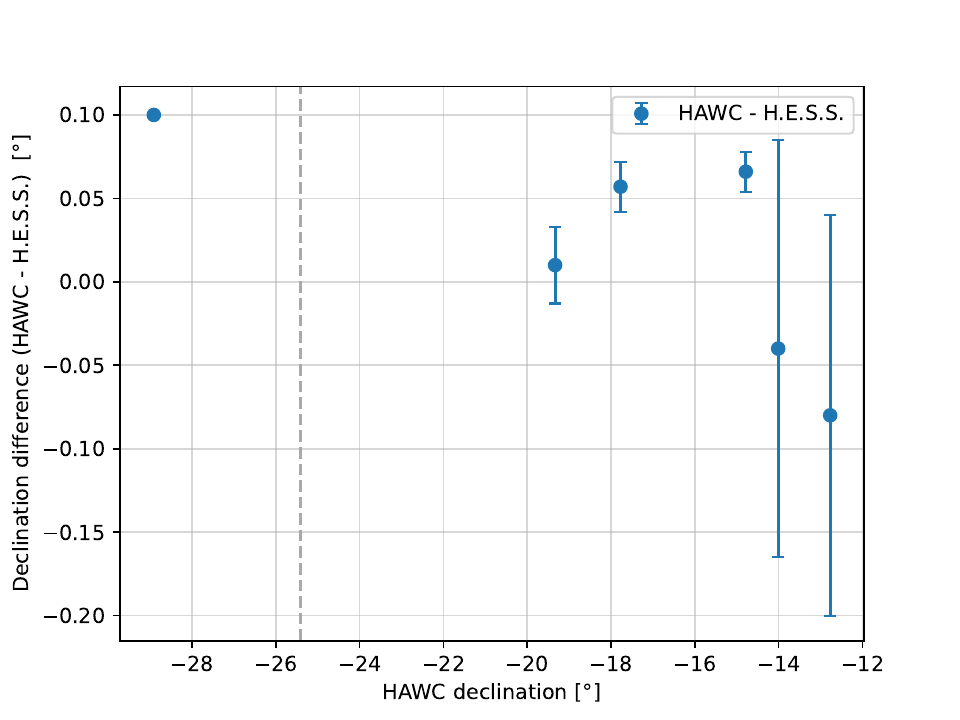}}
\caption{The measured declination of HAWC sources is compared to the TeV counterpart measurements from the IACT experiment H.E.S.S.; the vertical dashed line represents the declination of V4641~Sgr. The HAWC measurements generally agree with the source locations measured by H.E.S.S. within the provided uncertainties, covering a wide range of declinations including extreme values. The error bars in the plot represent the uncertainties derived from the H.E.S.S. results in the HGPS paper\cite{hgps2018hess}. 
}
    
\label{fig:decdiff}
\end{center}
\end{figure}   

We also studied the pointing accuracy. In the 3HWC catalog paper~\cite{albert20203hwc}, we quote the absolute pointing uncertainty of HAWC as 0.15$^\circ$ for sources at $-10^\circ$ or $+$$50^\circ$ declination, and 0.3$^\circ$ for sources at declination $-20^\circ$ or $+$$60^\circ$. With a better reconstruction algorithm, we have improved the pointing uncertainty. To validate this improvement, we conducted a series of tests.

We compare the measured declination of HAWC sources with the declination of their potential TeV counterparts as determined by H.E.S.S., chosen as a representative imaging atmospheric Cherenkov telescope (IACT). The source locations and their uncertainties were obtained from the H.E.S.S. Galactic plane survey (HGPS) paper \cite{hgps2018hess}. Extended Data Figure~\ref{fig:decdiff} shows the comparison between the declination of sources detected by HAWC and H.E.S.S., revealing substantial agreement between the two experiments for declinations ranging from $-10^\circ$ to $-30^\circ$. The error bars in the plot represent the uncertainties derived from H.E.S.S., which may be attributed to variations in individual source observation times. This comparison confirms the consistency and reliability of HAWC's pointing. 

\subsection{Source associations}
\label{sec:associations}
We searched for possible associates to the gamma-ray excess in high-energy catalogs. Unlike in radio and X-ray, a chance superposition of extragalactic background sources can be easily ruled out at 200~TeV. This is because most photons at this energy will be attenuated due to pair production with background radiation fields when traveling across the Milky Way and over extragalactic distances. At TeV energies, we found no known source within $3^\circ$ of the HAWC point sources. At GeV energies, three sources from the 4FGL catalog  \cite{4fgl_dr3_2022incremental} are found within the $0.5^\circ$ radius of the HAWC point sources, but all of them are outside of the $4.5\sigma$ excess region (see Figure~\ref{fig:map}). One of the three sources is a distant galaxy and the other two have no known associations. At X-ray wavelengths, we searched for counterparts in the Galactic X-ray Catalog from the NASA HEASARC database (\url{https://heasarc.gsfc.nasa.gov/W3Browse/all/xray.html}) and several X-ray binary catalogs, including the High-Mass X-ray Binaries Catalog (HMXBCAT)\cite{hmxbcat}, the Low-Mass X-Ray Binaries Catalog (LMXBCAT)\cite{lmxbc}, the Ritter Low-Mass X-Ray Binaries Catalog (RITTERLMXB)\cite{ritterlmxb}, the X-Ray Binaries Catalog (XRBCAT)\cite{xrbcat}, and the Integral IBIS 9-Year Galactic Hard X-ray Survey (INTIBISGAL)\cite{intibisgal}. Only a handful of sources are found in the vicinity, including IGR J18170-2511, XTE J1818-245, AX J1824.5-2451 (J1824-2452), and SAX J1810.8-2609. Three Galactic sources are located within $0.5^{\circ}$ of the northern and southern sources, namely, V4641~Sgr, IGR 1819.3-2542, and SWIFT J1818.7-2553. The latter two are known to be associated with V4641~Sgr\cite{bird2016ibis,oh2018105}.

\begin{table}[htpb]
\renewcommand{\arraystretch}{2}
\centering
\begin{tabular}{c c}
\hline
\hline
Data  selected according to MAXI & Number of Days \\ \hline
\begin{tabular}[c]{@{}c@{}}Oct 4--7 Nov 2021\\ (MJD 59491--59525)\end{tabular}        & 35 \\ \hline
\begin{tabular}[c]{@{}c@{}}Jan 12--Apr 1, 2020\\  (MJD 58860--58940)\end{tabular} & 80 \\ \hline
\begin{tabular}[c]{@{}c@{}}Aug 30--Oct 9, 2018\\  (MJD 58360--58400)\end{tabular}      & 40 \\ \hline
\begin{tabular}[c]{@{}c@{}}Jul 22--Aug 16, 2015\\  (MJD 57225--57250)\end{tabular}    & 25 \\ \hline
Total days                       & 180            \\ \hline
\hline
\end{tabular}
\caption{Data selected for the X-ray outbursts according to the Bayesian analysis of the daily light curves from the MAXI public data}
\label{tab:x-ray-burst}
\end{table}

\subsection{Time-dependent study}
\label{sec:selectxray}
V4641 Sgr is well known for its violent X-ray outbursts. 
To study the time dependency of the VHE gamma-ray emission, we carried out two distinct tests. Firstly, we split the data into two halves. The first half of data runs from November 2014 to May 2018, and the second half from May 2018 to June 2022; the significance and location of the gamma-ray excess do not show significant differences between the two halves. Secondly, we selected approximately 180 days of data during four recorded X-ray bursts from the Monitor of All-sky X-ray Image (MAXI) telescope \cite{matsuoka2009maxi}. We used the online Bayesian Blocks analysis tool on the MAXI website to select approximately 180 days of data during which V4641~Sgr was in an X-ray outburst state (\url{http://maxi.riken.jp/mxondem/}).
We searched for excess gamma-ray emission based on this combined dataset. Table~\ref{tab:x-ray-burst} details the dates chosen for the four confirmed outbursts. No significant excess is observed in the data for the selected time period.  The X-ray emission detected by Chandra during the 2020 outburst \cite{shaw2022high} exhibited significantly higher flux compared to the consistent gamma-ray flux observed by HAWC. Additionally, there was no noticeable increase in gamma-ray emissions during the periods corresponding to the X-ray outbursts. This disparity indicates that the gamma-ray and X-ray emission are not due to the same particle population. 
In summary, the HAWC-measured radiation shows no variation over different time periods, including recorded X-ray outbursts.

\subsection{Multi-wavelength observation of V4641~Sgr}

V4641~Sgr is famous for its super-Eddington X-ray outbursts. 
Its most dramatic flare happened in September, 1999 when the flux level reached 12.2~Crab, corresponding to a 1--10~keV luminosity of $(3$--$4)\times 10^{39}~ \mathrm{erg/s}$ \cite{revnivtsev2002super}. Some of the X-ray outbursts present lower luminosity. One explanation for the lower flux is that the central engine is potentially obscured by an optically thick outflow \cite{revnivtsev2002super, koljonen2020obscured, shaw2022high}, in which case the intrinsic X-ray luminosity was higher than observed.

H.E.S.S. conducted contemporaneous, VHE observations of V4641~Sgr with the Rossi X-ray Timing Explorer (RXTE) in 2018, with an overall livetime of 1.76\,h, yielding a non-detection in gamma rays \cite{abdalla2018search}. They reported an upper limit of the integral flux above 240~GeV as $I(E>240~\text{GeV})< 4.5 \times 10^{-12}~\mathrm{cm}^{-2}\,\mathrm{s}^{-1}$, which is consistent with HAWC's observation of the integral flux of $I(E>10~\text{TeV}) = 2.60^{+0.26}_{-0.05} \times 10^{-12}~\mathrm{cm}^{-2}\,\mathrm{s}^{-1}$. 
We note that the H.E.S.S. analysis only considered a single potential point source at the location of the binary, whereas the upper limit from HAWC is determined using the best-fit differential flux under the assumption of two point sources.

Radio observations by the Very Large Array (VLA) of the September, 1999 outburst events of V4641~Sgr resolved a bright, jet-like radio structure \cite{hjellming2000light}. These radio observations suggest that V4641~Sgr might be a Galactic ``microblazar," which means that its jet inclination is small or even aligned with the line of sight~\cite{orosz2001black,chaty2003optical,gallo2013v4641}.

\begin{figure}[tpb]
\begin{center}
\resizebox{0.8\textwidth}{!}{
\includegraphics[width=0.7\textwidth]{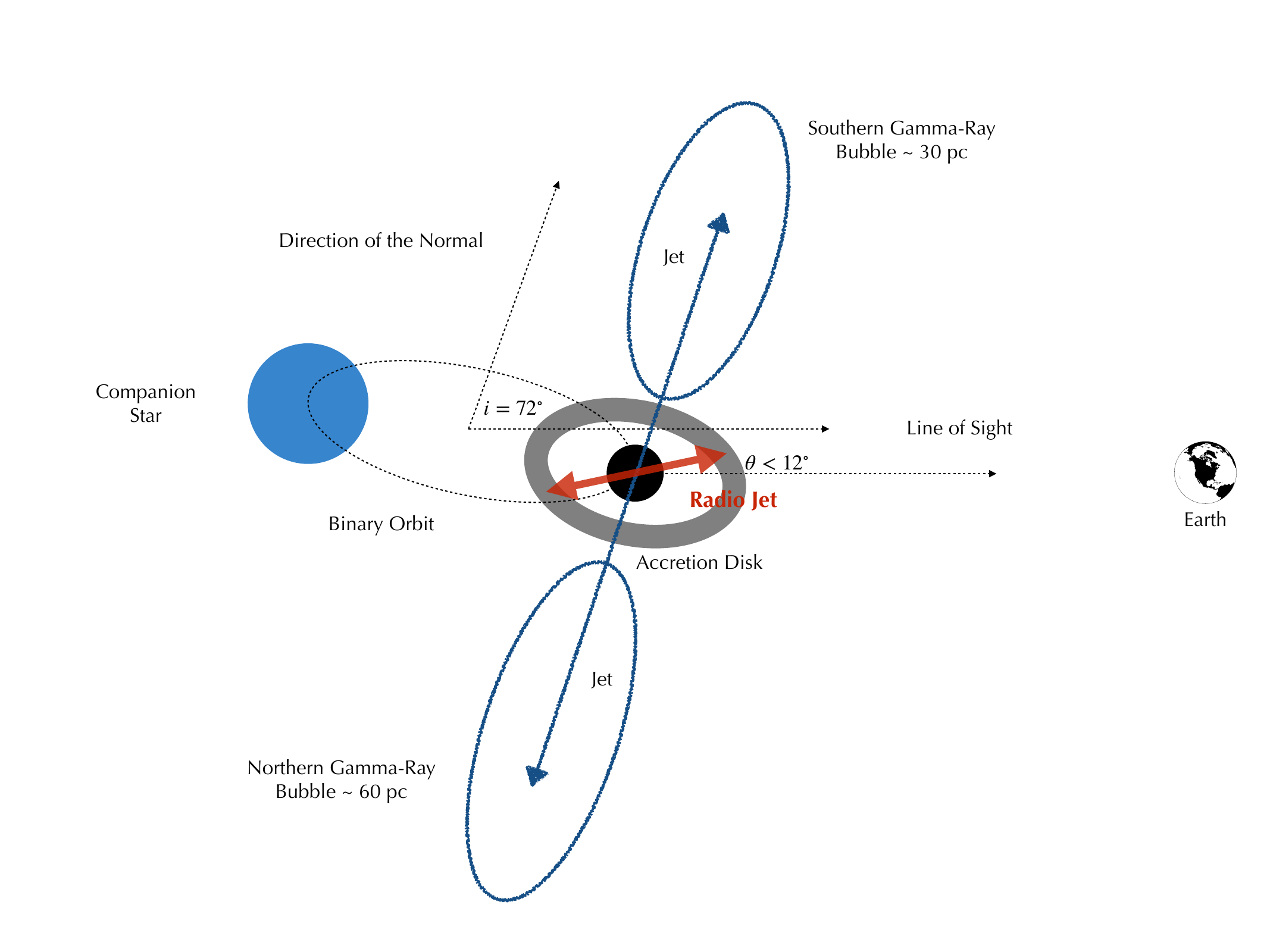}
}
\caption{ 
Schematic illustration of V4641~Sgr. 
Red arrows indicate the radio- and X-ray-emitting jets with sizes comparable to the scale of the binary system. They were detected during previous outbursts of this source with a jet inclination angle $\theta \lesssim 12^\circ$ ~\cite{orosz2001black}. Blue arrows indicate a set of gamma-ray-emitting jets suggested by this work, extending approximately 100~pc and oriented perpendicular to the accretion disk.
}

\label{fig:sketch}
\end{center}
\end{figure}

Figure~\ref{fig:sketch} presents a sketch of the binary system.  
The inclination angle, $i$, of V4641~Sgr, defined as the direction of the normal to the binary orbital plane with respect to the observer’s line of sight, is $\sim$$72.3^\circ \pm 4.1^\circ $ \cite{MacDonald:2014gpa}. In a standard system, where the disk is aligned with the orbital plane and the jets are perpendicular to the disk, this angle would be similar to the jet angle, $\theta$, defined as the angle of the jets with respect to the line of sight. It is not yet understood why the transient radio jets of V4641~Sgr present an unusual $\theta < 12^\circ$  \cite{orosz2001black}. The point-like gamma-ray sources, with an upper limit on the source extension of 0.2$^\circ$ at a 95\% confidence level resulting from the HAWC point-spread function (PSF) of 0.2$^\circ$ at energies greater than 30~TeV, would align with a scenario involving jets that are perpendicular to the accretion disk.

The absence of detectable persistent X-ray emission associated with the gamma-ray lobes could be due to either the low surface brightness of the X-ray counterpart,  which emits synchrotron radiation from secondary electrons, a lack of deep X-ray observations of the large-scale jets, or a combination of both factors.


\subsection{Gas distribution}
\label{sec:gas}
In the hadronic scenario, high-energy
protons interact with the ambient gas, resulting in the production of gamma rays through the decay process $\pi^{0}\rightarrow\gamma\gamma$. The molecular hydrogen survey \cite{2001ApJ...547..792D,Dame_2022} and the HI4PI 21-cm line emission survey of atomic hydrogen \cite{gas_kalberla2005, gas_maitra2006x, gas_pahari2015constraining} provide the gas column density from the direction of the V4641~Sgr region.

In order to obtain a realistic estimate of the gas density in the region of V4641~Sgr, we need to integrate the gas survey cubes over the range of velocities corresponding to the distance of V4641~Sgr and then divide by the velocity interval. The V4641~Sgr distance of 6.6~kpc roughly corresponds to a velocity of 90--100~km/s according to the Galactic rotation velocity curve \cite{clemens1985massachusetts}. Integrating the HI spectra over a range of velocities between 70~km/s and 120~km/s (this spread in velocities includes for instance possible internal turbulent motions of the gas), the first momenta yield $\mathrm{N}_H = 7.4 \times 10^{19}~\mathrm{cm}^{-2}$ and
$\mathrm{N}_H = 3.9 \times 10^{19}~\mathrm{cm}^{-2}$, for the southern and northern lobes, respectively. We also examined the molecular gas distribution at the locations of the two HAWC lobes by integrating the molecular hydrogen cubes over the velocity range 70--120~km/s \cite{2001ApJ...547..792D,Dame_2022}. As far as the molecular hydrogen is concerned, the maps contain high levels of noise
such that we cannot establish any emission from hadrons colliding off molecular hydrogen targets.

Assuming a diameter of $l\sim40$~pc for each lobe from an upper limit of 0.4$^\circ$ at 95\% confidence level for the Gaussian width of the sources, the gas density, $n$, is $n = 1~\mathrm{cm}^{-3}$ and $n = 0.5~\mathrm{cm}^{-3}$, for the south and north lobes, respectively.

\subsection{Very-high-energy emission due to hadronic interactions}
\label{sec:hadronic_scenario}

In the hadronic scenario, the interaction of high-energy protons with the surrounding gas in the source leads to the production of gamma rays through the decay process $\pi^{0}\rightarrow\gamma\gamma$. The acceleration of protons can occur at the termination shock, where the jets collide with the ISM, or along the jet itself, allowing them to be transported to the lobe region. An upper limit on the diameter of the two lobes in the two-source model, $l \sim 40$~pc, is
used to calculate the escape time.
If we account for particle escape, the energy budget channeled in the accelerated proton population depends upon the ratio between the particle (pp) cooling time and the escape time from the two lobes, $t_{pp}/t_{\, \rm esc}$. The pp cooling time is defined as $t_{pp} = 1/ (n \, \sigma_{pp} \, c)  \sim 1.6 \times {10}^{15} {(n/1\, \rm cm^{-3})}^{-1}~\mathrm{s}$, where $\sigma_{pp}$ is the cross section for pp interactions \cite{2009MNRAS.396.1629G}. 
The escape time,  $t_{\rm esc} \sim l^2/(2 \, D(E))$, where $l$ is the source size, depends upon the transport regime, $D(E)$. If we assume the diffusion at 1~PeV proceeds as inferred at GeV energies from the Galactic secondary ratio, with $D(1 \, {\rm PeV}) = \eta D_0 \sim \eta \, (3 \times 10^{30})~\mathrm{cm}^{2}/\mathrm{s}$, then the escape time may be described as $t_{\rm esc}\sim 3 \times 10^9 / \eta / (D_0/ (3 \times 10^{30}) \, \text{cm}^2/\text{s})~\mathrm{s}$. Under this scenario, the energy budget required in accelerated protons above 1~PeV (1~PeV protons produce roughly 100--200~TeV photons) would be around $\sim$$100\%$ of the Eddington luminosity,

\begin{equation}
   \dot{W_p} (E_p > 1 \, \text{PeV}) = L_{\gamma} \frac{t_{pp}}{t_{\text{esc}}} \simeq 10^{39} \, \eta \left(\frac{L_{\gamma}}{10^{34} \, \text{erg/s}}\right) \left(\frac{D_0}{ 3 \times 10^{30} \, \text{cm}^2/\text{s}}\right) \left(\frac{n}{1 \, \text{cm}^{-3}}\right)^{-1} \, \text{erg/s}\, ,
\end{equation}
where $n$ is the ambient gas density and $L_{\gamma}=9.13-9.50 \times 10^{33}$~erg/s for the north and south lobe, respectively. Transport could proceed in a slower ($\eta<1$) or faster way ($\eta>1$) with respect to the Galactic diffusion, $D_0$, derived from cosmic-ray secondary ratio at GeV energies. Cosmic-ray transport within acceleration sites might also proceed much more slowly--- in the extreme Bohm diffusion regime, $t_{\rm esc}\simeq 4 \times 10^3 \, {(E/1 \, {\rm PeV})}^{-1}~\mathrm{yr}$, the energy needed to be channeled into protons would be a small fraction of the Eddington luminosity.

\subsection{Very-high-energy emission due to leptonic interactions}

In the leptonic scenario, when assuming that the continuously accelerated electrons follow a differential-power-law spectrum extending up to at least 200~TeV, the northern and southern gamma-ray sources may be produced by electrons with an energy budget of
${7} \times {10}^{46}~\mathrm{erg}$ above 20~TeV in total.
The fraction of the energy of electrons released in inverse Compton scattering and not lost to synchrotron cooling is determined by the ratio of the energy density of the magnetic field (proportional to $B^2$) to the energy density of target photons, such that the expected luminosity in hard X-rays from the HAWC lobes should be $ L_X (> \text{few keV})\sim L_{TeV} (> 1~\text{TeV})  \,(\frac{B}{3 \rm \mu G})^{2} \sim 3-4 \times {10}^{34}\,(\frac{B}{3 \rm \mu G})^{2}~\mathrm{erg/s}$. We note that here we assume that the energy density of target photons is dominated by the cosmic microwave background (CMB) photon field even if in a small region close to the binary this is not true.

\subsection{Data availability}
The datasets analyzed during this study are available at a public repository maintained
by the HAWC Collaboration: \url{https://data.hawc-observatory.org/}

\subsection{Code availability.}
The study was carried out using the Multi-Mission Maximum Likelihood (3ML) HAWC Accelerated Likelihood (HAL) framework developed by the HAWC Collaboration. The software is open-source and publicly available
on Github: \url{https://github.com/threeML/hawc_hal}. The code distribution includes instructions on installation and usage.

\newpage

{\bf References}
\vspace{1em}


\end{document}